\documentclass[aps,onecolumn,preprintnumbers,superscriptaddress,nofootinbib]{revtex4}
\usepackage{graphicx}
\usepackage{amssymb}
\usepackage{color}
\usepackage{enumerate}

\newcommand{\be}{\begin{eqnarray}}
\newcommand{\ee}{\end{eqnarray}}

\newcommand{\beaa}{\begin{equation} \begin{array}{ll}}
\newcommand{\eeaa}{\end{equation} 	\end{array} }

\usepackage{graphicx}
\usepackage{mathrsfs}
\usepackage{float}
\usepackage{multirow}
\usepackage{booktabs}

\usepackage{float}

\begin{document}

\title{Approaching a spacetime singularity in conformal gravity}

\author{Leonardo Modesto}
\email{lmodesto@sustech.edu.cn}

\author{
Hui-Yu
Zhu}
\email{jacquelynzhy@gmail.com}

\author{
Jun-Yan 
Zhang}
\email{skycattmll@gmail.com}

\affiliation{Department of Physics, Southern University of Science and Technology, Shenzhen 518055, China}

\begin{abstract}
We hereby show that the Kasner spacetime turns out to be singularity-free in Einstein's conformal gravity in vacuum or in presence of matter. Such a statement is based on the regularity of the curvature invariants and on the geodesic completion of the spacetime when it is probed by massive, massless, and conformally coupled particles. 
As a universal feature of the regular metric, nothing can reach the singularity located in $t=0$ in a finite amount of proper time (for massive particles and conformally coupled particles) or for finite values of the affine parameter (for massless particles).

\end{abstract}

\maketitle

\section{Introduction}

One of the main issues in General Relativity is the presence of spacetime singularities in most of the solutions of the Einstein's equation of  motion (EoM) that play a crucial role in cosmology and astrophysics. We can remind here very well known examples like the FRW spacetime, the Schwarzschild metric, the Kerr metric, etc. One could think that a local or nonlocal extension of Einstein's theory can smooth out all the singularities. However, in general, this is not the case as proved in \cite{Li:2015bqa}, and in order to cure the problem of spacetime singularities we need something more than engineering the Lagrangian describing the gravitational interaction. 
We are thus tempted to look for a possible solution to the problem of singularities in a principle of symmetry present at the fundamental level, but somehow broken at the macroscopic scale.

Going back up to 1977 \cite{NK}, records show that Narlikar and Kembhavi proposed and proved that the Weyl's conformal symmetry could be the correct fundamental symmetry principle within which the singularities could be tamed. 

Let us now be more quantitative.  
 The Einstein's conformal invariant theory is specified by the following action functional,
\be
 S=\int \! d^4x \sqrt{-\hat{g}} \, \left(\phi^2\hat{R}+6\hat{g}^{\mu\nu}\partial_\mu\phi \partial_\nu \phi \right)
 \label{CEG}
 \, ,
\ee
where $\hat{g}_{\mu\nu}$ is the spacetime metric and $\phi$ is a scalar field (the dilaton). Since the presence of the dilaton, the theory is not only invariant under general coordinate transformations, but also under the following conformal 
transformation:
\be 
\hat{g}^\prime_{\mu\nu}=\Omega^2\hat{g}_{\mu\nu} \, ,  \quad \phi^\prime=\Omega^{-1}\phi \, .
\label{WR}
\ee
The Einstein-Hilbert gravity (defined below in (\ref{EH})) is recovered when the Weyl conformal invariance is broken spontaneously in exact analogy with the Higgs mechanism in the standard model of particle physics (for more details we refer the reader to \cite{FiniteConformal,Bambi:2016wdn}). 

One possible vacuum of the theory (\ref{CEG}) (exact solution of the EoM) is 
$\phi = {\rm const.} = 1/\sqrt{16 \pi G}$, together with $R_{\mu\nu} =0$. 
Therefore, making the replacement $\phi = 1/\sqrt{16 \pi G} + \varphi$ in the action and using the conformal invariance to eliminate the gauge dependent degree of freedom $\varphi$, we finally get the Einstein-Hilbert action in absence of matter, 
\be
S_{\rm EH} = \frac{1}{16 \pi G} \int d^4 x \! \sqrt{-\hat{g}} \, \hat{R} \, .
\label{EH}
\ee
Hence, Einstein's gravity is simply the theory (\ref{CEG}) in the spontaneously broken phase of conformal invariance \cite{Bambi:2016wdn}. The generalization to gravity coupled to matter is straightforward: it is sufficient to properly introduce the dilaton in order to make the whole action conformal invariant.

Besides the constant vacuum, if a metric $\hat{g}_{\mu\nu}$ is an exact solution of the EoM, thus it is also the rescaled spacetime with a non trivial profile for the dilaton field, namely 
 \be
 \hat{g}_{\mu\nu}^*=S(r) \hat{g}_{\mu\nu} \, \quad \phi^* = S(r)^{-1/2} \, \phi . 
 \label{rescaling} 
 \ee
Therefore, we can use the above rescaling to construct other exact and singularity-free solutions of the theory. Assuming that the Weyl conformal symmetry is spontaneously broken, the regular spacetimes are those selected by Nature.

In this paper we address the general case of a singular Kasner \cite{Kasner} spacetime. Indeed, as proven in \cite{BKL}, it is the most general way of approaching a time-like singularity.

In $D=4$ the Kasner metric is an homogeneous, but not isotropic, exact solution of Einstein's gravity in vacuum (we here assume a constant value for the dilaton field, hence, the action is (\ref{EH})), and it reads \cite{Kasner}:
\be
   { \rm{d}} \hat{s}^2= \hat{g}_{\mu\nu}{\rm{d}}x^\mu{\rm{d}}x^\nu=-{\rm{d}}t^2+a_1(t){\rm{d}}x_1^2+a_2(t){\rm{d}}x^2_2+a_3(t){\rm{d}}x^2_3,
\ee
where the factors $a_i(t)$ are:
\be 
a_i(t)=t^{2p_i} \,  \quad {\rm and} \quad \sum_{i=1}^3 p_i=\sum_{i=1}^3 p_i^2=1 \, ,
\label{ai}
\ee
and the latter two conditions come from the EoM. 
We can equivalently parametrize the numbers $p_i$ above in terms of a single independent parameter $u\geq1$ \cite{BKL}, namely 
\be
    p_1=-\frac{u}{u^2+u+1} \, , 
    \quad
    p_2=\frac{u+1}{u^2+u+1} \, , 
    \quad 
    p_3=\frac{u(u+1)}{u^2+u+1} \, .
    \label{p123}
\ee
The ranges for the parameters $p_i$ are:
\be
    -\frac{1}{3}\le p_1\le0 \, , \quad 0\le p_2\le\frac{2}{3} \, , \quad \frac{2}{3}\le p_3\le1 \, .
\ee 

The Kasner spacetime is singular in $t=0$ and not geodetically complete because any particle massive or massless can reach the singularity in a finite amount of proper time. However, as stated in (\ref{rescaling}), a general rescaling of a solution (in our case the Ricci flat Kasner spacetime) is also an exact solution.
In other words, the following metric and the dilaton, 
\be
\hat{g}^*_{\mu\nu} = S(t) \hat{g}_{\mu\nu}^{\rm Kasner} \, , \qquad \phi^* = S(t)^{-1/2} \, \phi^{\rm Kasner} \, ,
\label{KR}
\ee
provide a candidate singularity-free spacetime metric for a proper choice of the rescaling factor $S(r)$. 

In the rest of the paper we will prove that the spacetime described by the metric (\ref{KR}), for two particular choices of $S(r)$, has no curvature singularities, and, most importantly, it is geodetically complete. As explained above, the singularity-free background is one of the possible vacua obtained breaking spontaneously the Weyl conformal symmetry. Therefore, the regularity of the spacetime has to be tested looking at curvature scalars invariant only respect to general coordinate transformations. 
Obviously there are many space-times that share the same properties, hence, it will be Nature to select one of them through the mechanism of spontaneous symmetry breaking. However, we will see that all the regular spacetimes share the same universal property, namely the former singularity in $t=0$ is unattainable.

\section{Avoiding the Kasner curvature singularity in conformal gravity}
Starting from eq.(\ref{KR}) in the previous section, we show that a proper choice of the rescaling $S(t)$ removes the curvature singularity in $t=0$. Since the Kasner spacetime is Ricci flat, before the rescaling the first non-trivial curvature invariant is the Kretschmann scalar, namely

\be
     K := R_{\alpha \beta \gamma \delta} R^{\alpha \beta \gamma \delta} = C_{\alpha \beta \gamma \delta} C^{\alpha \beta \gamma \delta}, 
   \label{Kre}
\ee
where in the last equality we used that $R_{\alpha \beta}=0$ and introduced the Weyl tensor $C_{\alpha \beta \gamma \delta}$. Under the Weyl rescaling (\ref{WR}) the Weyl tensor is invariant, namely 
\be
C^{\prime \alpha}\,_{ \beta \gamma \delta} = C^{\alpha}\,_{ \beta \gamma \delta}\, .
\ee
Hence, the Kretschmann scalar (\ref{Kre}) turns into 
\be
 K^\prime = C^{\prime \alpha}\,_{\beta \gamma \delta} \, C^{\prime \mu}\,_{\nu\rho\sigma} \,  g^\prime_{\alpha \mu} g^{\prime \beta \nu} g^{\prime \gamma \rho} g^{\prime \delta \sigma} 
 = C^{\alpha}\,_{\beta \gamma \delta} \, C^{\mu}\,_{\nu\rho\sigma} \,  g_{\alpha \mu} g^{\beta \nu} g^{\gamma \rho} g^{\delta \sigma} S(t) \, S^{-1}(t) \, S^{-1}(t)\,  S^{-1}(t) = \frac{K}{S^2(t)} . 
 \label{K)}
\ee
Expressing $a_i$ in (\ref{ai}) in terms of the parameter $u\geqslant1$, 
the scalar (\ref{K)}) reads:
\begin{equation}
    K'=\frac{16u^2(1+u)^2}{t^4(1+u+u^2)^3 \, S^2(t)}.
\end{equation}
The Kretschmann scalar for the Kasner metric is recovered when we take $S(t)=1$. Indeed, if $S(t)=1$ the curvature invariant blows up in $t=0$. However, by choosing a proper rescaling factor $S(t)$, the curvature singularity can be avoided. An appropriate and minimal choice of $S(t)^{-1}$ should be at least quadratic in $t$ near $t=0$, e.g.
\be
S=\frac{t^2+B^2}{t^2}, 
\label{SSS}
\ee
where $B \neq 0$ is a constant with length dimension. 

Actually, we can push further our analysis of the spacetime regularity.  Since the Kasner metric is Ricci flat, the most general nonzero local curvature invariant is made of only the Weyl tensor for the metric $\hat{g}_{\mu\nu}$. Assuming to build up a scalar contracting $n$ Weyl tensors we get the following curvature invariant,
\be
{\bf C}^n := C^{\mu_1}\,_{\nu_1 \rho_1 \sigma_1} C^{\mu_2}\,_{\nu_2 \rho_2 \sigma_2}  \dots 
C^{\mu_i}\,_{\nu_i \rho_i \sigma_i}
\dots 
C^{\mu_n}\,_{\nu_n \rho_n \sigma_n} \, g_{\mu_1 \mu_2} \,  g^{\nu_1 \nu_2} \dots g_{\mu_i \mu_n} ,
\ee
which undergoes the following rescaling,
\be
{\bf C}^{\prime n} = S^{-n}(t) \, {\bf C}^n \, ,
\ee
is regular in $t=0$.

Since now $K$ and any general curvature invariant is finite, for any value of $t$ the curvature singularity is removed. 
However, to claim that the metric is singularity free, we have to prove that the spacetime is geodetically complete. Indeed, the Taub-Nut's metric is a famous counterexample of singularity-free spacetime, but not geodetically complete.

\section{Geodesic completion}\label{geo}
In this section we will address the issue of the geodesic completion of the rescaled Kasner spacetime (\ref{KR}) studying the propagation of  massive, massless, and conformally coupled particles (\ref{KR}).

\subsection{A massive particle probe} \label{massiveparticle}
In order to prove the geodesic completion of the spacetime we here consider the motion of a massive particle towards the point $t=0$. The action for a massive particle in the spacetime (\ref{KR}) is:
\be
S_{m} = - m \int \sqrt{ - \hat{g}_{\mu\nu} d x^\mu d x^\nu} 
=  - m \int \sqrt{ - \hat{g}_{\mu\nu} \frac{d x^\mu}{d \lambda} \frac{d x^\nu}{d \lambda} } 
\, d \lambda \equiv \int d \lambda \, \mathcal{L}_m \, . 
\label{Spc}
\ee
Notice that respect to (\ref{KR}) we redefined $\hat{g}^*_{\mu\nu} \rightarrow \hat{g}_{\mu\nu}$ to avoid a cumbersome notation. 
Since the Lagrangian does not depend on the coordinates $x_i$ ($i=1,2,3$) we have three Killing vectors corresponding to three conserved quantities, i.e. 
\be
\frac{\partial \mathcal{L}_m}{\partial \dot{x}^i} = - \frac{m^2 \hat{g}_{ii} \dot{x}^i}{ \mathcal{L}_m} = {\rm const.} = e_i \, 
\quad \Longrightarrow \quad 
\dot{x}^i = - \frac{e_i \, \mathcal{L}_m}{m^2 \hat{g}_{ii}} ,
\label{xdot}
\ee
where in the first formula on the left we do not have to sum on the repeated index $i$.
Since we are interested in evaluating the proper time need to a massive particle to reach the point $t=0$, we must choose the proper time gauge, namely $d \lambda^2 = d \tau^2$. Hence, the Lagrangian simplifies to $\mathcal{L}_m = - m$ and $\dot{x}^i = - e_i/m \hat{g}_{ii}$. From $ds^2/d \tau^2 = -1$ and (\ref{xdot})  we get
\be
\hat{g}_{00} \dot{t}^2 + \hat{g}_{11} \dot{x}^2 + \hat{g}_{22} \dot{x}^2 + \hat{g}_{33} \dot{x}^2 = -1 \, ,
\label{xdot2}
\ee
which turns into 
\be
\hat{g}_{00}  \dot{t}^2 + \frac{e_1^2}{\hat{g}_{11}} +   \frac{e_2^2}{\hat{g}_{22}}  +  \frac{e_3^2}{\hat{g}_{33}} = -1 \qquad 
\Longrightarrow \qquad 
S(t) \dot{t}^2 =  \frac{e_1^2}{S(t) t^{2 p_1}} +   \frac{e_2^2}{S(t) t^{2 p_2}}  +  \frac{e_3^2}{S(t) t^{2 p_3}} + 1
\, ,
\label{imply}
\ee
when (\ref{xdot}) and (\ref{KR}) are replaced in (\ref{xdot2}). 

Finally, the proper time is obtained by  integrating (\ref{imply}), namely 
\be
\tau = - \int \frac{t^2+B^2}{\sqrt{t^2 \left(e_1^2 \, t^{\frac{2 (u+1)^2}{u^2+u+1}}+e_2^2 \,  t^{\frac{2 u^2}{u^2+u+1}}+e_3^2 \, 
   t^{\frac{2}{u^2+u+1}}+t^2 + B^2\right)  }} dt 
   \label{intM}
\ee
where the minus sign on the right side comes from integrating from bigger to smaller values of $t$ for the case of a particle that is approaching the singularity. 
Since we are interested in the regime $t \rightarrow 0$, we can expand (\ref{intM}) for small values of $t$ gaining the following simple integral (notice that the three exponents of the $t$ variable inside the round brackets are all positive), 
\be
\tau \simeq  - \int \frac{B}{t} dt  = -B\log t \, , 
\label{inteM0}
\ee
which goes to infinity for $t\rightarrow 0$. Therefore, a massive particle can never reach the singularity in $t=0$ and, thus, the spacetime is geodetically complete. 
It deserves to be notice that the result (\ref{inteM0}) is independent on the value of parameter $u$ and the constants $e_i$, basically because all the exponents of $t$ in (\ref{intM}) are positive, while for $e_i=0$ the constant $B$ dominates inside the round brackets. 

\subsection{A massless particle probe}\label{massless1}
We now study the propagation of massless particles (e.g. photons) in the Kasner spacetime. 
For massless particles the correct action, which is invariant under reparametrizations of the world line, $p^\prime= f(p)$, is: 
\be
S_{\gamma} = \int \mathcal{L}_{ \gamma } d \lambda = \int e(p)^{-1} \phi^2\hat{g}_{\mu\nu}  \frac{d x^\mu}{d p} \frac{d x^\nu}{d p} d p \, 
\label{Lm0}
\ee
where $e(p)$ is an auxiliary field that transforms as $e^\prime(p^\prime)^{-1} = e(p)^{-1} (d p^\prime/d p)$ in order to guarantee the invariance of the action. The action (\ref{Lm0}) is not only invariant under general coordinate transformations, but also under the Weyl conformal rescaling (\ref{WR}). 

The variation respect to $e$ gives:
\be
\frac{ \delta S_{\gamma}}{\delta e} = 0 \quad \Longrightarrow \quad  - \int d p  \, \frac{\delta e}{e^2} \phi^2 \, \hat{g}_{\mu\nu} \, \dot{x}^{\mu} \dot{x}^\nu = 0 \quad \Longrightarrow \quad   d \hat{s}^2 =  \hat{g}_{\mu\nu} dx^\mu d x^\nu = 0 \, ,
\label{ds0}
\ee
consistently with the main property of massless particles.

The variation respect to $x^\mu$ gives the geodesic equation in presence of the dilaton field, namely 
\be
\frac{D^2(\hat{g}) x^\lambda}{d p^2} + 2 \frac{\partial_\mu \phi}{\phi} \frac{ d x^\mu}{d p} \frac{ d x^\lambda }{d p} 
-  \frac{\partial^\lambda \phi }{\phi} \frac{ d x^\mu}{d p} \frac{ d x_\mu }{d p} 
= 0 \, ,
\label{geom0}
\ee
where $D^2(\hat{g})$ is the covariant derivative respect to the metric $\hat{g}_{\mu\nu}$.

However, when we contract equation (\ref{geom0}) with the velocity $d x_\lambda/d p$ and we use $d\hat{s}^2 =0$ obtained in (\ref{ds0}), we get the following on-shell condition, 
\be
\frac{d x_\lambda}{d p} \, \frac{D^2(\hat{g}) x^\lambda}{d p^2} + 2  \frac{d x_\lambda}{d p} \, \frac{\partial_\mu \phi}{\phi} \frac{ d x^\mu}{d p} \frac{ d x^\lambda }{d p} 
-  \frac{d x_\lambda}{d p} \, \frac{\partial^\lambda \phi }{\phi} \frac{ d x^\mu}{d p} \frac{ d x_\mu }{d p} 
= 0 \quad \Longrightarrow \quad \frac{d x_\lambda}{d p} \, \frac{D^2(\hat{g}) x^\lambda}{d p^2} = 0 \, .
\ee

Therefore, the $\frac{D^2(\hat{g}) x^\lambda}{d p^2}$ must be proportional to the velocity, namely 
\be
\frac{D^2(\hat{g}) x^\lambda}{d p^2} = f \, \frac{d x^\lambda}{d p}  \quad (f = {\rm const.}) 
\label{constre}
\ee
because the velocity is null on the light cone. 
Under a reparametrization of the world line $q = q(p)$ eq.(\ref{constre}) becomes 
\be
\frac{d^2 x^{\lambda} }{dq^2} + \Gamma^\lambda_{\mu\nu}  \frac{d x^\mu }{d p}  \frac{d x^\nu }{d p} 
=  \frac{d x^\lambda}{d p} \left( \frac{dp}{dq} \right) \left(f \frac{dq}{dp} - \frac{d^2 q}{dp^2} \right) .
\label{parame}
\ee
Choosing the dependence of $q$ on $p$ such us to make vanish the right-hand side of (\ref{parame}), we end up we the geodesic equation in the affine parametrization. Hence, we can redefine $q \rightarrow \lambda$ and, finally, we get the affinely parametrized geodesic equation for photons in the metric $\hat{g}_{\mu\nu}$, 
\be
\frac{D^2(\hat{g}) x^\lambda}{d \lambda^2} = 0 \, .
\label{affine}
\ee

We can naw investigate the conservations laws based on the symmetries of the metric. Let us consider the following scalar,
\be
\hat{\alpha} = \hat{g}_{\mu\nu} v^\mu \frac{d x^\nu}{d \lambda} \, .
\label{alpha}
\ee
where $v^\mu$ is a general vector. Taking the derivative of (\ref{alpha}) respect to $\lambda$ and using the geodesic equation (\ref{affine}) we get

\be
\frac{d}{d \lambda} \hat{\alpha} = \frac{1}{2}  v^\mu \partial_\mu \hat{g}_{\rho \nu} \frac{d x^\rho}{d \lambda} 
\frac{d x^\nu}{d \lambda} + g_{\mu\nu} \partial_\rho v^\mu \frac{d x^\nu}{d \lambda} \frac{d x^\rho}{d \lambda} 
= \frac{1}{2} [\mathcal{L}_v \hat{g} ]_{\rho \nu }  \frac{d x^\rho}{d \lambda} \frac{d x^\nu}{d \lambda} \, , 
\ee
where $[\mathcal{L}_v \hat{g}]$ is the Lie derivative of $\hat{g}_{\mu\nu}$ by a vector field $v^\mu$. 
Thus, if $v^\mu$ is a Killing vector field, namely $[\mathcal{L}_v \hat{g}]=0$, $\hat{\alpha}$ is conserved:
\be
\frac{d}{d \lambda} \left[ \hat{g}_{\mu\nu} v^\mu \frac{d x^\nu}{d \lambda} \right] = 0 
\, . 
\label{CL}
\ee

For the case of the Kasner metric we have three Killing vectors, namely 
\be    
    v_1=(0, \, 1, \, 0, \, 0) \, , \quad 
    v_2=(0, \,  0, \,  1, \, 0) \, , \quad 
    v_3=(0, \,  0, \,  0, \, 1) \, , 
 \label{killing}
\ee
that when replaced in (\ref{CL}) give the following conserved quantities,
\be
&& e_1 = \hat{g}_{\alpha \beta} \, v_1^\alpha \dot{x}^\beta = \hat{g}_{11} \, \dot{x}^1 \, , \nonumber \\
&& e_2 = \hat{g}_{\alpha \beta} \, v_2^\alpha \dot{x}^\beta =  \hat{g}_{22}  \, \dot{x}^2 \, ,  \nonumber \\
&& e_2 = \hat{g}_{\alpha \beta} \, v_2^\alpha \dot{x}^\beta =  \hat{g}_{33} \,  \dot{x}^3 \, . 
\label{e123}
\ee

From (\ref{ds0}) or $d\hat{s}^2=0$ it follows 
\be
d \hat{s}^2 = - \dot{t}^2 + t^{2 p_1} \, \dot{x}_1^2 + t^{2 p_2} \, \dot{x}_2^2 + t^{2 p_3} \, \dot{x}_3^2 = 0 \, .
\label{explids}
\ee
Now we extract  $\dot{x}_1$, $\dot{x}_2$, and $\dot{x}_3$ from (\ref{e123}) and we replace the results in (\ref{explids}) to finally we end up with the following equation,
\be
&&  \dot{t}^2 = t^{2 p_1} \left( \frac{e_1}{g_{11}} \right)^2 
+ t^{2 p_2} \left( \frac{e_2}{g_{22}} \right)^2 
+ t^{2 p_3} \left( \frac{e_3}{g_{33}} \right)^2 \, ,  \nonumber \\
&& \dot{t}^2 =  \frac{e_1^2}{S(t)^2 t^{2 p_1}}
+  \frac{e_2^2}{S(t)^2 t^{2 p_2}} 
+  \frac{e_3^2}{S(t)^2 t^{2 p_3}} \, .
\label{finalEq0}
\ee

Integrating (\ref{finalEq0}) we get the following result for the affine parameter as a function of $t$, 
\be
\lambda = - \int \frac{B^2+t^2}{\sqrt{t^4 \left(e_1^2 \, t^{-2 p_1}+e_2^2 \, t^{-2 p_2}+e_3^2 \, t^{-2 p_3}\right)}} dt 
\ee
According to the definitions (\ref{p123}), the latter integral can be expressed in terms of the parameter $u$, namely 
\be
\lambda = - \int\frac{B^2+t^2}{\sqrt{t^4 \left(e_1^2 \, t^{\frac{2 u}{u^2+u+1}}+e_2^2 \, t^{-\frac{2 (u+1)}{u^2+u+1}}+e_3^2 \,    t^{-\frac{2u(1+u)}{u^2+u+1}}\right)}} dt \, . 
\label{intgamma2}
\ee
In order to investigate the geodesic completion of the spacetime we can expand (\ref{intgamma2}) near $t=0$ where, for $e_3 \neq 0$, the leading term in the round brackets is the last one for $u>1$, while the exponents of the last two terms are equal for $u=1$.   
Hence, for $u >1$ the following simple integral determines the affine parameter near the Kasner singularity,
\be
\lambda \simeq - \int\frac{B^2}{e_3 \,  t^{{2-\frac{u (u+1)}{u^2+u+1}}}} dt = \frac{B^2 \left(u^2+u+1\right) }{|e_3| } \frac{1}{t^{\frac{1}{u^2+u+1}}}
\, . 
\ee

For $e_3 =0$ and $u>1$, we can again forget the first term in the round brackets and (\ref{intgamma2}) turns into:
\be
\lambda  \simeq - \int\frac{B^2}{|e_2| \,  t^{{2-\frac{u+1}{u^2+u+1}}}} dt = \frac{B^2 \left(u^2+u+1\right) }{|e_2|  \, u^2} \frac{1}{t^{\frac{u^2}{u^2+u+1}}}
\, . 
\ee

For $u=1$ we have the following three cases: $e_2\neq 0$, $e_3=0$, or $e_2 =  0$, $e_3 \neq 0$, or 
 $e_2 \neq 0$, $e_3 \neq 0$, but the solutions have the same scaling, namely 
\be
\lambda  \simeq \frac{3 B^2 }{|e_3| t^{1/3}} \quad {\rm or} \quad \lambda  \simeq \frac{3 B^2 }{|e_2| t^{1/3}} \quad {\rm or} \quad \lambda \simeq \frac{3 B^2 }{ \sqrt{e_2^2+e_3^2} \, t^{1/3}} \, .
\ee

The last case is when $e_2 = e_3 =0$, then
\be
\lambda \simeq - \int   \frac{B^2 }{|e_1| \, t^{\frac{u^2}{u^2+u+1}}} dt = \frac{B^2 \left(u^2+u+1\right)}{|e_1| (1+u^2)} \frac{1}{ t^{\frac{1+u^2}{u^2+u+1}}}
\, . 
\ee

Since $u$ is positive, all the above solutions for the affine parameter $\lambda$ approach $+ \infty$ for $t\rightarrow0$. Therefore, massless particles can never reach $t=0$ for finite values of the affine parameter $\lambda$.

\subsection{A conformally coupled particle probe}
For the last we would like to use a conformally coupled partile to probe the spacetime near $t=0$. The action compatible with Weyl conformal invariance reads:
\be
S_{\rm c} = - \int \sqrt{ - \phi^2  \hat{g}_{\mu\nu} d x^\mu d x^\nu} 
=  - \int \sqrt{ - \phi^2 \hat{g}_{\mu\nu} \frac{d x^\mu}{d \lambda} \frac{d x^\nu}{d \lambda} } 
\, d \lambda \equiv \int d \lambda \, \mathcal{L}_{\rm c} \, . 
\label{SCC}
\ee
 Now we can repeat the same steps of section (\ref{massiveparticle}), but notice that the mass in the action has been replaced with the dilaton field.

As for the case of the massive particle the Lagrangian does not depend on the coordinates $x_i$ ($i=1,2,3$) we  have three Killing vectors and three conserved quantities again, i.e.
\be
\frac{\partial \mathcal{L}_m}{\partial \dot{x}^i} = - \frac{\phi^2 \hat{g}_{ii} \dot{x}^i}{ \mathcal{L}_m} = {\rm const.} = e_i \, 
\quad \Longrightarrow \quad 
\dot{x}^i = - \frac{e_i \, \mathcal{L}_{\rm c}}{\phi^2 \hat{g}_{ii}} ,
\label{xdot0}
\ee
where again we do not have to sum on the repeated index $i$.
We are interested in evaluating the proper time need to a conformally copled particle to reach the point $t=0$, hence, we must choose the proper time gauge, namely $d \lambda^2 = d \tau^2$, and the Lagrangian simplifies to 
\be
\mathcal{L}_{\rm c} = - \phi \quad {\rm and} \quad  \dot{x}^i = - \frac{e_i}{\phi \, \hat{g}_{ii}} \, .
\label{xdot1}
\ee
 From $ds^2/d \tau^2 = -1$ 
 we get
\be
\hat g_{00} \dot{t}^2 + \hat{g}_{11} \dot{x}^2 + \hat{g}_{22} \dot{x}^2 + \hat{g}_{33} \dot{x}^2 = -1 \, ,
\label{line}
\ee
which turns into 
\be
\hat g_{00}  \dot{t}^2 + \frac{e_1^2}{\phi^2 \hat{g}_{11}} +   \frac{e_2^2}{\phi^2 \hat{g}_{22}}  +  \frac{e_3^2}{ \phi^2 \hat{g}_{33}} = -1 \qquad 
\Longrightarrow \qquad 
S(t) \dot{t}^2 =  \frac{e_1^2}{ t^{2 p_1}} +   \frac{e_2^2}{ t^{2 p_2}}  +  \frac{e_3^2}{ t^{2 p_3}} + 1
\, ,
\label{implyconf}
\ee
when (\ref{xdot1}), (\ref{KR}), and (\ref{ai}) are replaced into (\ref{line}). 

The vacuum solution with rescaling (\ref{SSS}) is not geodetically complete. Indeed, the reader can easily check that a conformally couple particle moving in a Kasner spacetime (\ref{KR}) with rescaling (\ref{SSS}) can reach $t=0$ in a finite amount of proper time. Thus we consider the following rescaling factor, 
\be
S(t) = \left(\frac{B^2 + t^2}{t^2} \right)^2 .
\label{SSSS}
\ee

Finally, the proper time is obtained integrating (\ref{imply}), namely 
\be
\tau = - \int \sqrt{\frac{(B^2+t^2)^2}{t^4 \left(1+ e_1^2 \, t^{\frac{2 u}{u^2+u+1}+2}+e_2^2 \,  t^{- \frac{2  (u+1)}{u^2+u+1}}+e_3^2 \, t^{- \frac{2 u (u+1)}{u^2+u+1}} \right)} } dt.
   \label{intCP}
\ee

We now expand (\ref{intCP}) near $t=0$. For $e_3 \neq 0$, the leading term in the round brackets is the last one for $u>1$, while the exponents of the last two terms are equal for $u = 1$.  
Hence, for $u>1$ the following simple integral determines the proper time for a conformally coupled particle  near the Kasner singularity,
\be
\tau  \simeq - \int\frac{B^2}{|e_3| \,  t^{{2-\frac{u (u+1)}{u^2+u+1}}}} dt = \frac{B^2 \left(u^2+u+1\right) }{|e_3| } \frac{1}{t^{\frac{1}{u^2+u+1}}}
\, . 
\ee

For $e_3 =0$ and $u>1$, we can again forget the first term in the round brackets and (\ref{intgamma2}) turns into:
\be
\tau  \simeq - \int\frac{B^2}{|e_2| \,  t^{{2-\frac{u+1}{u^2+u+1}}}} dt = \frac{B^2 \left(u^2+u+1\right) }{|e_2|  \, u^2} \frac{1}{t^{\frac{u^2}{u^2+u+1}}} 
\, . 
\ee

For $u=1$ we have the following three cases as for the massless particles: $e_2\neq 0$, $e_3=0$, or $e_2 =  0$, $e_3 \neq 0$, or 
 $e_2 \neq 0$, $e_3 \neq 0$, but the solutions have the same scaling, namely 
\be
\tau \simeq \frac{3 B^2 }{|e_3| \, t^{1/3}} \quad {\rm or} \quad  \tau \simeq \frac{3 B^2 }{|e_2| \, t^{1/3}} \quad {\rm or} \quad  \tau  \simeq \frac{3 B^2 }{\sqrt{e_2^2+e_3^2} \, t^{1/3}} \, .
\ee

For the last we consider the case $e_2 = e_3 =0$, 
\be
\tau {\simeq} - \int   \frac{B^2 }{t^2} dt = \frac{B^2}{t} \,  .
\ee
Since $u$ is positive, in all cases the proper time $\tau$ approaches $+ \infty$ for $t\rightarrow0$. Therefore, also conformally coupled particles never reach $t=0$ in a finite amount of proper time for the proper rescaling (\ref{SSSS})\footnote{Making use of the coordinate transformation $dt'=\sqrt{r/2M}dr$, the Schwarzschild metric can be approximated by a special case of Kasner spacetime near the central singularity, namely for $u=1$. As a result, after the rescaling, we get a geodetically complete Schwarzschild spacetime.}.

\section{The Kasner spacetime in the presence of matter}

In the last section we want to show that the presence of matter does not change the outcomes of the previous sections.
For the sake of simplicity we here consider a scalar massless field without any potential term for which an analytic exact solution was derived in \cite{Scalar}. The Lagrangian for matter reads\footnote{The conformal invariant action is obtained replacing the scalar field $\Phi$ with $\Phi/\phi$ where $\phi$ is the dilaton.}:
\be
\mathcal{L}_\phi = - g^{\mu\nu} \partial_\mu \Phi\partial_\nu \Phi ,
\ee
while the EoM after selecting a constant vacuum for the dilaton $\phi$ are:
\be
R_{\mu\nu} = \nabla_\mu \Phi \nabla_\nu \Phi \, , \quad 
\Box \Phi = 0 \, .
\ee
The exact solution of the above EoM is:
\be
&& \Phi = q \, \log t \, , \quad -\sqrt{\frac23}\leq q \leq \sqrt{\frac23}\, ,  \nonumber \\
&& p_1+p_2 +p_3 = 1 \, , \quad p_1^2 + p_2^2 + p_3^2 = 1 - q^2 \, , 
\ee
and the ranges for the parameters $p_i$ are:
\be
    -\frac{1}{3}    \le p_1      \le \frac{1}{3} \, , \quad 0     \le p_2   \le    \frac{2}{3} \, , \quad        \frac{1}{3} \le p_3 \le 1 \, .
\ee

Similarly to the Kasner solution in vacuum, it is convenient to parametrize the exponents in the following form \cite{Scalar},
\be
 &&
    p_1 =\frac{-u}{1+u+u^2} \, , \nonumber \\
&&    p_2 =\frac{1+u}{1+u+u^2}\left\{u-\frac{u-1}{2}\left[1-\left(1-\beta^2\right)^{\frac{1}{2}}\right]\right\} \, , \nonumber \\
&&    p_3 =\frac{1+u}{1+u+u^2}\left\{1+\frac{u-1}{2}\left[1-\left(1-\beta^2\right)^{\frac{1}{2}}\right]\right\} \, , \nonumber \\
&&    \beta = \frac{ 2 \left(1+u+u^2\right)^2q^2}{\left(u^2-1\right)^2} \, . 
\ee
The invariance of the exponents $p_i$ under the transformation $u \rightarrow 1/u$ confines ourself to the examination of the region $-1 \leqslant u \leqslant 1$, and the region of admissible values of $u$ and $q$ is determined by the inequality \cite{Scalar}
\be
\beta^2 \leqslant 1 \, .
\ee
 Obviously, for $q=0$ we get back to the Kasner solution in vacuum.

\subsection{Singularity Problem}

The Kretschmann scalar in presence of matter depends also on the parameter $q$, i.e.  
\begin{equation}
    K=\frac{16u^2(1+u)^2+8qu(q+u+u^2)^2+3q^2(1+u+u^2)^3}{t^4(1+u+u^2)^3}\, , 
\end{equation}
which is singular for $t=0$.

The singularity free metric in conformal gravity is again (\ref{KR}) with scaling factor (\ref{SSS}), but also the scalar field undergoes a conformal rescaling, namely
\be
\Phi^{*}(t) = q \,  S^{-1/2}(t) \, \log t = q \, \log t \, \left( \frac{t^2}{t^2 + B^2}  \right)^{\frac{1}{2}} \, ,
\ee
which is regular when $t \rightarrow 0$. 

In order to check the curvature regularity of the spacetime we compute again the Kretschmann invariant,
\begin{equation}
    K=\frac{16u^2(1+u)^2+8qu(q+u+u^2)^2+3q^2(1+u+u^2)^3}{t^4(1+u+u^2)^3 \, S^2(t)}\, , 
\end{equation}
which is obviously regular because the presence of the parameter $q$ does not affect the behaviour of the scalar $K$ near $t=0$.

 \subsection{Geodesic Completion}
The geodesic completion can be proved on the same lines of the previous sections \ref{geo}. 

\subsubsection{A massive particle probe}

For massive particle, just like (\ref{intM}), we find the following integral for the proper time:
\begin{equation}
    \tau=-\int\frac{B^2+t^2}{\sqrt{t^2 \left(e_1^2 \, t^{\frac{2 u}{u^2+u+1}+2}+e_2^2 \, t^{\frac{\sqrt{1-\beta ^2}-\sqrt{1-\beta ^2} u^2+u^2+1}{u^2+u+1}}+e_3^2 \, t^{2-\frac{2 (u+1) \left[1-\frac{1}{2} \left(\sqrt{1-\beta ^2}-1\right) (u-1)\right]}{u^2+u+1}}+t^2+B^2\right)}}dt.
    \label{tauWM}
\end{equation}
The region of validity for the parameters $u$ and $\beta^2$ are $-1\leqslant u\leqslant1$,  $0\leqslant\beta^2\leqslant1$ and $-\sqrt{2/3}\leqslant q\leqslant\sqrt{2/3}$ respectively \cite{Scalar}, hence, the three exponents of the three terms of the $t$ variable in the round brackets are all positive.
Therefore, we can expand (\ref{tauWM}) for $t\to0$ and obtain the same result 
as in (\ref{inteM0}), namely 
\begin{equation}
    \tau  \simeq - \int \frac{B}{t} dt  = - B \log t \, ,
\end{equation}
which diverges for 
$t\to0$, and the spacetime is geodetically complete.

\subsubsection{A massless particle probe}
\begin{figure}
\begin{center}
\includegraphics[height=8cm]{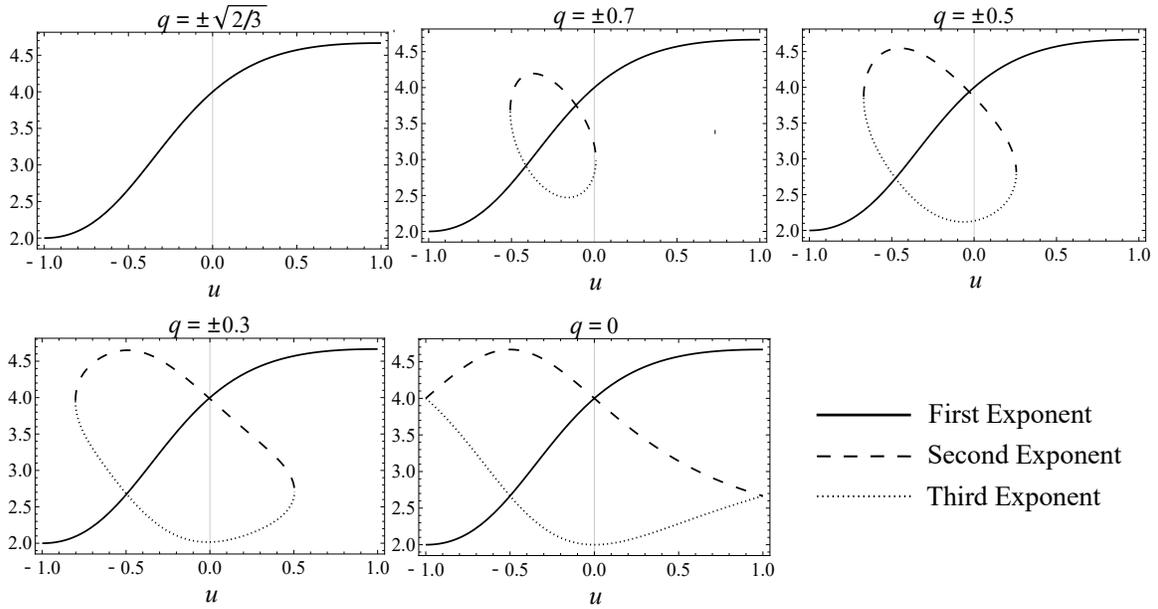}
\end{center}
\caption{Plots for the three exponents inside the square root in (\ref{mattermassless}) as functions of the parameter $u$ for five different values of $q$}
   \label{diagram}
\end{figure}

For massless particles, one can check that the factor (\ref{SSS}) makes the spacetime geodetically complete only in vacuum, namely $q=0$, which is exactly the case of section \ref{massless1}. Thus we here consider the factor (\ref{SSSS}) and we obtain the following relation between the affine parameter $\lambda$ and the time coordinate $t$ is: 
\begin{equation}
    \lambda= - \int \frac{B^2+t^2}{\sqrt{e_1^2 \, t^{4+\frac{2 u}{u^2+u+1}}+e_2^2 \, t^{4-\frac{2 (u+1) \left[\frac{1}{2} (u-1) \left(\sqrt{1-\frac{4 q^4 \left(u^2+u+1\right)^4}{\left(u^2-1\right)^4}}-1\right)+u\right]}{u^2+u+1}}+e_3^2 \, t^{4-\frac{2 (u+1) \left[1-\frac{1}{2} (u-1) \left(\sqrt{1-\frac{4 q^4 \left(u^2+u+1\right)^4}{\left(u^2-1\right)^4}}-1\right)\right]}{u^2+u+1}}}}{\rm{d}}t \, .
    \label{mattermassless}
\end{equation}
The three exponents of the variable $t$ inside the square root are now more involved because depend on two parameters, i.e. $u$ and $q$, hence, we make use of the plots in Fig.\ref{diagram}. 

%

\begin{figure}
\begin{center}
\includegraphics[height=5.5cm]{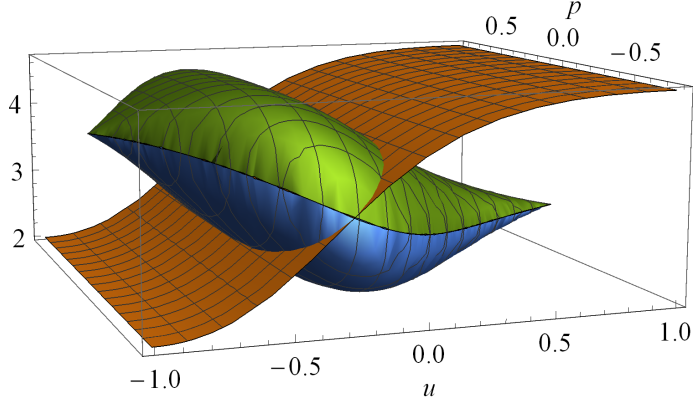}
\end{center}
\caption{$3D$-plot for the three exponents inside the square root in (\ref{mattermassless}) as functions of $u$ and $q$.}
 \label{3D}
\end{figure}

In Fig.\ref{diagram} we show only five cases, while in Fig.\ref{3D} all the cases are covered making use of a three-dimensional plot. 

%
Looking at the $3D$ plot one can see that 
for all values of the parameter $u$ in the range $-1\leqslant u\leqslant1$, and for all values of $q$,  
the leading term for $t\rightarrow 0$ would be either the first or the third one under the square root in (\ref{mattermassless}) whether $e_1\neq0$ or $e_3\neq0$.

If the first term in the variable $t$ in (\ref{mattermassless}) is the leading one (we should require $e_1\neq0$), the integral simplifies to: 
\begin{equation}
    \lambda\simeq\int-\frac{B^2}{ |e_1| \, t^{2+\frac{u}{u^2+u+1}}}{\rm{d}}t=\frac{B^2 \left(u^2+u+1\right) t^{-\frac{u}{u^2+u+1}-1}}{|e_1|(u+1)^2}.
    \label{mattermassless1}
\end{equation}
If the third term in $t$ in (\ref{mattermassless}) is dominant ($e_3\neq0$), we get: 
\begin{equation}
    \lambda\simeq
    -\frac{B^2 t^{\frac{(u+1) \left[1-\frac{1}{2} (u-1) \left(\sqrt{1-\frac{4 q^4 \left(u^2+u+1\right)^4}{\left(u^2-1\right)^4}}-1\right)\right]}{u^2+u+1}-1}}{|e_3| \left(\frac{(u+1) \left[1-\frac{1}{2} (u-1) \left(\sqrt{1-\frac{4 q^4 \left(u^2+u+1\right)^4}{\left(u^2-1\right)^4}}-1\right)\right]}{u^2+u+1}-1\right)}.
    \label{mattermassless3}
\end{equation}
Finally, if $e_1=e_3=0$, and $e_2\neq0$, we find:  
\begin{equation}
    \lambda\simeq-\frac{B^2 t^{\frac{(u+1) \left(\frac{1}{2} (u-1) \left(\sqrt{1-\frac{4 q^4 \left(u^2+u+1\right)^4}{\left(u^2-1\right)^4}}-1\right)+u\right)}{u^2+u+1}-1}}{|e_2| \left(\frac{(u+1) \left(\frac{1}{2} (u-1) \left(\sqrt{1-\frac{4 q^4 \left(u^2+u+1\right)^4}{\left(u^2-1\right)^4}}-1\right)+u\right)}{u^2+u+1}-1\right)}.
    \label{mattermassless2}
\end{equation}

Since the exponent of the time coordinate $t$ in the three solutions above is always negative, then, the affine parameter $\lambda$ approach $+\infty$ for $t\to0$ and the spacetime is geodetically complete.

 \subsubsection{A conformally coupled particle probe}

For a conformally coupled particle, using (\ref{SSSS}) we can get the following proper time, 
\begin{equation}
    \tau=-\int\frac{B^2+t^2}{t^2\sqrt{1+e_1^2 \, t^{\frac{2 u}{u^2+u+1}}+
    e_2^2 \, t^{-\frac{2 (u+1) \left[\frac{1}{2} (u-1) \left(\sqrt{1-\frac{4 q^4 \left(u^2+u+1\right)^4}{\left(u^2-1\right)^4}}-1\right)+u\right]}{u^2+u+1}}+
    e_3^2 \, t^{-\frac{2 (u+1) \left[1-\frac{1}{2} (u-1) \left(\sqrt{1-\frac{4 q^4 \left(u^2+u+1\right)^4}{\left(u^2-1\right)^4}}-1\right)\right]}{u^2+u+1}}}}dt \, .
\end{equation}

When the three exponents of the variable $t$ inside the square root are all positive, for $t\to 0$ the leading term at the denominator of the integrand is $t^2$ and we get:
\begin{equation}
    \tau\simeq-\int\frac{B^2}{t^2}dt= \frac{B^2}{t} \, . 
\end{equation}

On the other hand, when one among of the three exponents inside the square root is negative for $t\to 0$, we need to apply the same as for equation (\ref{mattermassless}). 

If the term proportional to $e_1\neq0$ is the leading one, the outcome of the integral reads: 
\begin{equation}
    \tau\simeq \frac{B^2 \left(u^2+u+1\right) t^{-\frac{u}{u^2+u+1}-1}}{|e_1|(u+1)^2} \, .
\end{equation}
If the term proportional to $e_3\neq0$ is dominant, we have
\begin{equation}
    \tau\simeq-\frac{B^2 t^{\frac{(u+1) \left[1-\frac{1}{2} (u-1) \left(\sqrt{1-\frac{4 q^4 \left(u^2+u+1\right)^4}{\left(u^2-1\right)^4}}-1\right)\right]}{u^2+u+1}-1}}{|e_3| \left(\frac{(u+1) \left[1-\frac{1}{2} (u-1) \left(\sqrt{1-\frac{4 q^4 \left(u^2+u+1\right)^4}{\left(u^2-1\right)^4}}-1\right)\right]}{u^2+u+1}-1\right)} \, .
\end{equation}
Finally, if $e_1=e_3=0$, but $e_2\neq0$, and the exponent of the term proportional to $e_2$ is negative, we have \begin{equation}
    \tau \simeq  -\frac{B^2 t^{\frac{(u+1) \left(\frac{1}{2} (u-1) \left(\sqrt{1-\frac{4 q^4 \left(u^2+u+1\right)^4}{\left(u^2-1\right)^4}}-1\right)+u\right)}{u^2+u+1}-1}}{e_2 \left(\frac{(u+1) \left(\frac{1}{2} (u-1) \left(\sqrt{1-\frac{4 q^4 \left(u^2+u+1\right)^4}{\left(u^2-1\right)^4}}-1\right)+u\right)}{u^2+u+1}-1\right)} \, .
\end{equation}

Since $-1\leqslant u \leqslant 1$ and $-\sqrt{2/3} \leqslant q \leqslant \sqrt{2/3}$, all possible forms of the proper time $\tau$ above approach to $+\infty$ for $t\to0$. Therefore, conformally coupled particle never approach $t=0$ in a finite amount of proper time in presence of matter.

\section{Conclusions} 

On the footprints of previous works \cite{NK, Bambi:2016wdn, FiniteConformal}, we have explicitly proved that in a conformal invariant theory the Kasner spacetime, in presence or not of matter, is singularity free. Indeed, in any conformal invariant theory, we have a large number of vacuum solutions at our disposal all connected by a Weyl transformation. Hence, assuming that the conformal symmetry is spontaneously broken towards a regular spacetime, we end up with a class of geodetically complete manifolds. We have explicitly proved that the Kretschmann invariant is finite everywhere in spacetime, and most importantly we have shown that the point $t=0$ is never reached by any massive, massless, or conformally coupled particle. The latter property is general regardless of the ambiguity in the choice of the vacuum. 
Indeed, in conformal gravity it turns out the universal feature that singularities are unattainable points of the spacetime \cite{Chakrabarty:2017ysw}.

\end{document}